\documentclass{article}
\usepackage{graphicx}
\usepackage{xparse}
\usepackage{amsmath}
\usepackage{color}
\usepackage[numbers]{natbib}

\usepackage{authblk}

\usepackage[font=footnotesize]{caption}

\usepackage[utf8]{inputenc}
\usepackage{amsmath}
\usepackage{physics}
\usepackage{amssymb}
\usepackage{mathtools}
\usepackage[version=4]{mhchem}
\usepackage{graphicx}
\usepackage{xcolor}
\usepackage[section]{placeins}
\usepackage[left=3.5cm,right=3.5cm,top=3cm, bottom=3cm]{geometry}
\usepackage{tikz}
\usetikzlibrary{shapes.geometric, arrows}
\usepackage{booktabs}
\usepackage{subcaption}
\usepackage[flushleft]{threeparttable}

\usepackage{authblk}
\usepackage{hyperref}
\hypersetup{
    colorlinks,
    linkcolor={red!70!black},
    citecolor={blue},
    urlcolor={blue!40!black}
  }
\numberwithin{equation}{section}

\title{How to use quantum computers \\ for biomolecular free energies}

\author[1]{Jakob G\"unther}
\author[2]{Thomas Weymuth}
\author[2]{Moritz Bensberg}
\author[1,3]{Freek Witteveen}
\author[4,5]{Matthew S. Teynor}
\author[6]{F. Emil Thomasen}
\author[7,8]{Valentina Sora}
\author[4]{William Bro-J{\o}rgensen}
\author[2]{Raphael T. Husistein}
\author[2]{Mihael Erakovic}
\author[1]{Marek Miller}
\author[9]{Leah Weisburn}
\author[9]{Minsik Cho}
\author[2]{Marco Eckhoff}
\author[10]{Aram W. Harrow\thanks{corresponding author: aram@mit.edu}}
\author[7,8]{Anders Krogh\thanks{corresponding author: akrogh@di.ku.dk}}
\author[9]{Troy Van Voorhis\thanks{corresponding author: tvan@mit.edu}}
\author[6]{Kresten Lindorff-Larsen\thanks{corresponding author: lindorff@bio.ku.dk}}
\author[4,5]{Gemma Solomon\thanks{corresponding author: solomon@chem.ku.dk}}
\author[2]{Markus Reiher\thanks{corresponding author: mreiher@ethz.ch}}
\author[1]{Matthias Christandl\thanks{corresponding author: christandl@math.ku.dk}}

\affil[1]{University of Copenhagen, Dept. of Mathematical Sciences, Quantum for Life Centre, Copenhagen, Denmark}
\affil[2]{ETH Zurich, Dept. of Chemistry and Applied Biosciences, Zurich, Switzerland.}
\affil[3]{QuSoft, Centrum voor Wiskunde en Informatica, Amsterdam, The Netherlands.}
\affil[4]{University of Copenhagen, Dept. of Chemistry and Nano-Science Center, Copenhagen, Denmark.}
\affil[5]{University of Copenhagen, Niels Bohr Institute, NNF Quantum Computing Progr., Copenhagen, Denmark.}
\affil[6]{University of Copenhagen, Dept. of Biology, Linderstr{\o}m-Lang Centre for Protein Science, Copenhagen, Denmark.}
\affil[7]{University of Copenhagen, Dept. of Computer Science, Quantum for Life Centre, Copenhagen, Denmark.}
\affil[8]{University of Copenhagen, Dept. of Public Health, Copenhagen, Denmark.}
\affil[9]{Massachusetts Institute of Technology, Dept. of Chemistry, Cambridge, MA, USA.}
\affil[10]{Massachusetts Institute of Technology, Center for Theoretical Physics, Cambridge, MA, USA.}

\date{  \vspace*{-1em} \small June 24, 2025 }

\begin{document}

\maketitle

\vspace{-0.8cm}
 
\begin{abstract}
    Free energy calculations are at the heart of physics-based analyses of biochemical processes. They allow us to quantify
    molecular recognition mechanisms, which  determine a wide range of biological phenomena from how cells send and receive signals to how pharmaceutical compounds can be used to treat diseases. Quantitative and predictive free energy calculations require computational models that accurately capture both the varied and intricate electronic interactions between molecules as well as the entropic contributions from motions of these molecules and their aqueous environment.
   However, accurate quantum-mechanical energies and forces can only be obtained for small atomistic models, not for large biomacromolecules. Here, we demonstrate how to consistently link accurate quantum-mechanical data obtained for substructures to the overall potential energy of biomolecular complexes by machine learning in an integrated algorithm. We do so using a two-fold quantum embedding strategy where the innermost quantum cores are treated at a very high level of accuracy. We demonstrate the viability of this approach for the molecular recognition of a ruthenium-based anticancer drug by its protein target, applying traditional quantum chemical methods. As such methods scale unfavorable with system size, we analyze requirements for quantum computers to provide highly accurate energies that impact the resulting free energies. Once the requirements are met, our computational pipeline \mbox{FreeQuantum} is able to make efficient use of the quantum computed energies, thereby enabling quantum computing enhanced modeling of biochemical processes. This approach combines the exponential speedups of quantum computers for simulating interacting electrons with modern classical simulation techniques that incorporate machine learning to model large molecules.
\end{abstract}

\section{Introduction}
\sloppy

Free energy calculations represent the state of the art for predicting the affinity between biomolecules in aqueous solution. These methods provide an atomistic understanding of molecular recognition and ligand binding \cite{wang2015accurate,mobley2017predicting,abel2017advancing}. This understanding is key to deciphering the mechanisms that underlie processes in cells with far-reaching consequences for biological function and hence for disease and health. For these reasons, free energy calculations are increasingly implemented as part of strategies for discovering and improving small molecule modulators of biological function \cite{cournia2021free,wade2022alchemical}.

\fussy 

For biomolecular recognition driven by the change in free energy, the prototypical example is the interaction of a ligand (such as a small-molecule drug) with a biomacromolecular target, typically a protein. The ability to predict such free energy differences at high accuracy and speed---and for a wide range of types of molecules---would enable a revolutionary ability to describe biological processes at the molecular level, to engineer proteins for predefined
functions, and to improve treatment of diseases via improved drugs. The accurate calculation of the free energy difference of such a binding process, however, requires one to deal with the enormous size of the configuration space that needs to be sampled for a proper representation of the thermodynamical ensemble. For each sampled atomistic structure, furthermore, an accurate estimate for the interaction energy needs to be computed.

While the sampling is a computational problem that in principle can be dealt with by well-established approaches (e.g., \cite{lelievre2010free,york}), the interaction energy computation would be more accurate when based on a quantum-mechanical description of the electrons, in particular for complex molecular systems which are difficult to handle with standard force fields. Machine learning-based potentials provides a potential route to bridge the accuracy of quantum-chemical calculations with the sampling needed for accurate biological free-energy calculations. Previous work in this area has, for example, used various flavours of such potentials---typically trained on density functional theory (DFT) calculations across a broad set of fragment molecules---to perform free-energy calculations \cite{SabanesZariquiey2025QuantumBind,Crha2025Alchemical,Semelak2025Advancing,Wang2025AccurateFreeEnergy,Wang2025DesignSpace}.

Accurate quantum-mechanical calculations carried out by traditional computers, however, scale highly unfavorably with system size and are therefore almost always out of reach for routine studies of ligand-binding. The basic reason for this can be found in the exponential overhead in the memory requirement of the description of a quantum-mechanical system on a traditional, classical computer, even if only the protein-ligand interface is treated in a fully quantum-mechanical way. A future quantum computer, by contrast, is in principle able to represent the entire wavefunction, since the number of qubits required scales linearly with the number electrons, opening the potential to overcome this `curse of dimensionality' \cite{Feynman1982,lloyd,baiardi_quantum_2023}. 

To realize this potential, energy-sampling methods must leverage highly accurate quantum-mechanical calculations; however, current computational pipelines are not 
set up to handle such high-level methods --- neither for calculations that can be carried out with traditional quantum chemical methods today, nor for those carried out with future quantum computers. Without this capability, the promise of quantum computing for broad application in biochemistry remains unclear.

\begin{figure}
    \centering
    \includegraphics[width=0.4\linewidth]{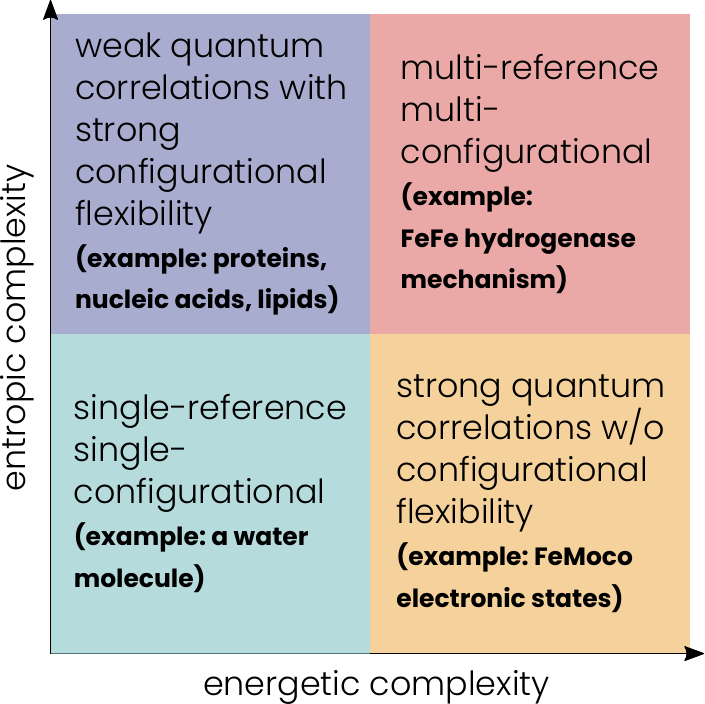}
    \caption{Biomolecular quantum simulation quadrangle: The complexity of the electronic structure problem is indicated on the horizontal axis. The importance and complexity of the sampling problem is indicated on the vertical axis. Considering the four paradigmatic situations where the energetic and entropic complexities are low or high gives rise to four quadrants. In each, we indicate a characteristic example. 
}
    \label{fig:parameterspace-overview}
\end{figure}

We note that in different biomolecular situations, there are different levels of importance of the energetic contributions when compared to the entropic contributions. In Fig.~\ref{fig:parameterspace-overview} we sketch this 'biomolecular quantum simulation quadrangle' with low versus high complexity of energy when compared to entropy, and give key examples.
The situation with a low entropic complexity ranges from simple closed-shell molecules (such as a single water molecule), which can be tackled with traditional methods, to open-shell $3d$ transition metal clusters (such as the FeMo-cofactor of nitrogenase) for which traditional methods struggle but quantum computing is promising \cite{ReiherPNAS}. The situation with high entropic complexity ranges from the case of comparatively weak electronic correlations (including most biochemical situations) to the case where high electronic complexity meets large configurational freedom (e.g., the FeFe hydrogenase which involves an iron-sulfur cluster as active site whose mechanism depends on the configurational flexibility of the surrounding protein scaffold (see, e.g., \cite{Finkelmann,Happe2024Protein, Stiebritz2012Hydrogenases})). 
We note that for many systems in the top-left corner of the quadrangle standard force fields can in principle perform well, especially if the system only has top-row elements from the periodic table. However, in the presence of transition metals, accuracy is often poor.

In this work, we present a framework and implementation of an end-to-end pipeline for the calculation of free energies that makes efficient use of expensive high-accuracy quantum-mechanical calculations. We demonstrate the accuracy, reliability and plasticity of our algorithmic workflow with the example of a ruthenium-based anticancer compound binding to a target protein (GRP78/NKP-1339), where the quantum data is obtained by traditional quantum chemical wave function-based methods. Our pipeline has the flexibility to allow for the direct replacement of this traditional computing engine with a quantum computing. We provide requirements for when this replacement will impact the free energy calculation. This is an important step towards charting and realizing the conditions for quantum advantage in computing free energies for large and complex biological systems.

In previous work \cite{Q4Bio-ML,Q4Bio-Embedding}, we have already demonstrated that (i) we can efficiently represent a hybrid low-cost-quantum--classical model of a protein--guest complex by a machine learning potential from which a free energy of binding can be obtained and that (ii) this machine learning model can be improved by transfer learning with local high-accuracy quantum energies obtained in a second embedding, where smaller quantum cores embedded in the large, low-cost quantum region of the hybrid model. 

\begin{figure}
    \centering
    \includegraphics[width=\textwidth]{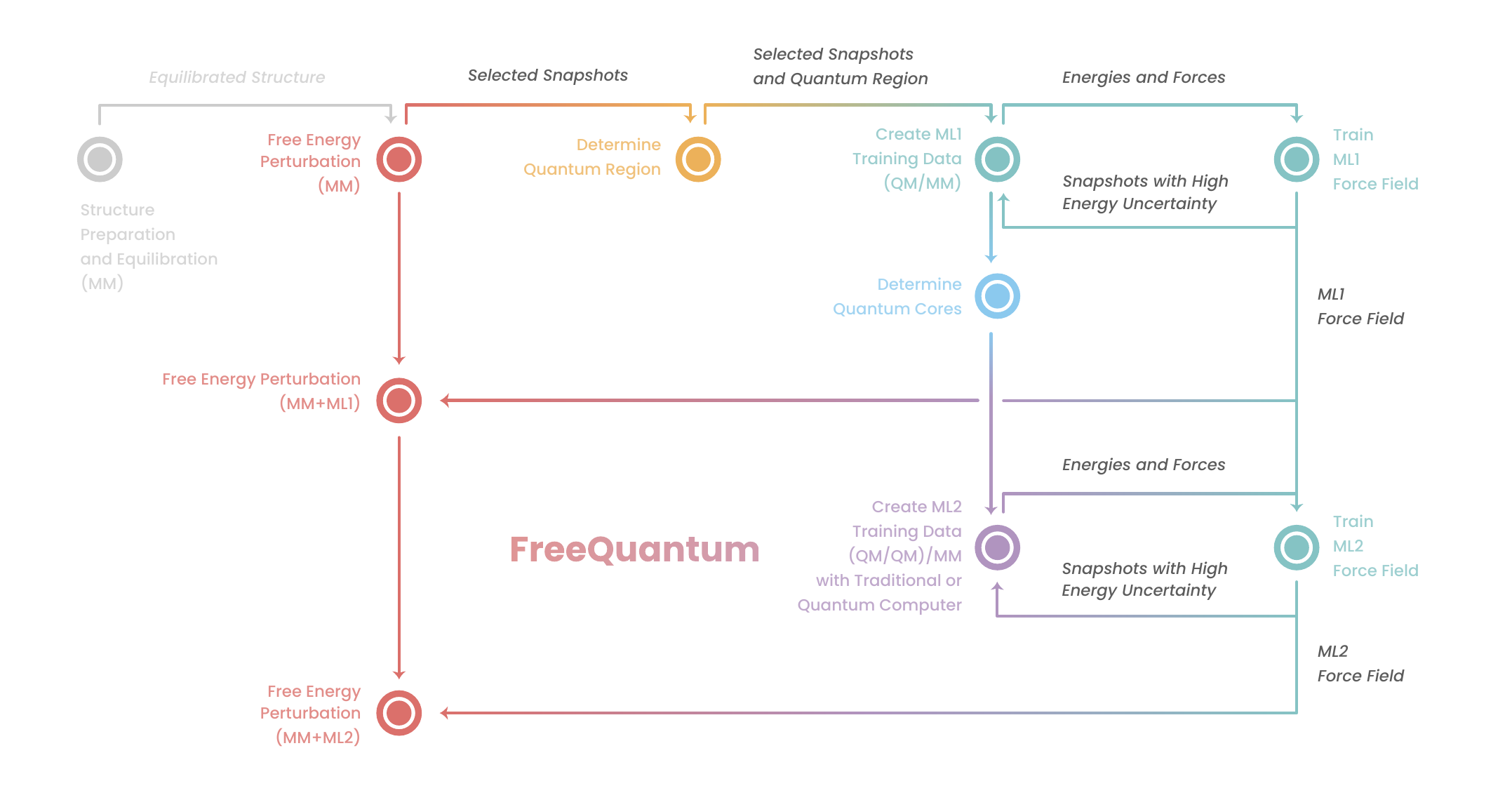}
    \caption{Workflow of the FreeQuantum pipeline: After a structure preparation step for the host protein and bound guest molecule by equilibration (grey), structural sampling with classical force fields starts in an initial alchemical Free Energy Perturbation (FEP) step (red, top).  
    The structures are then forwarded to the hybrid QM/MM modeling step, which defines the quantum region that covers the parts of the protein-ligand system to be subjected to a quantum-mechanical description (orange). For the resulting QM/MM structures with associated energies and forces, a first machine learning (ML) potential (ML1) is trained with an active learning loop to interlace the QM subsystem data with the MM force field data (turquoise, top). The loop involves non-equilibrium (NEQ) switching calculations, the final result of which is output as the MM+ML1 FEP(red, middle). In a further step, the QM/MM hybrid model is refined by introducing a second type of embedding, 
   where quantum-in-quantum subregions (quantum cores) are defined (blue), for which then highly accurate quantum-mechanical calculations are carried out, either by traditional quantum chemical calculations or by future quantum computation (violet). The resulting QM/QM/MM data are fed into a refining process of the earlier ML model (ML1) by transfer learning (turquoise, bottom). Sampling from this refined model (ML2) yields the third tier of free energy results (red, bottom). 
    With the active learning and the FEP calculations, we automated all parts of the pipeline which otherwise would require frequent manual intervention and substantial human time. 
}
    \label{fig:pipeline}
\end{figure}

In this work, we now build the FreeQuantum pipeline by incorporating these building blocks into a general framework that (i) can deal with complex atomistic situations such as transition metal complexes, open-shell electronic structures and multi-configurational electronic substructures in the quantum cores, that (ii) can switch from traditional electronic structure methods for the quantum cores to quantum algorithms, 
and that (iii) can be driven fully automatically according to various outer constraints such as the amount of computational resources, classical and quantum, available (e.g., measured in terms of the number and quality of qubits available). In this way, the pipeline can tailor the atomistic modeling of the multilayer embedding procedure to the hardware available for the most accurate quantum calculations on the quantum cores. For a graphical representation of our FreeQuantum pipeline, see Fig.~\ref{fig:pipeline}.

We demonstrate the FreeQuantum pipeline by calculating the free energy of binding of the aforementioned open-shell transition-metal-containing anticancer drug to its protein target with a series of traditional correlated electronic structure methods. Based on our earlier work developing qubit-efficient quantum algorithms for groundstate energy estimation \cite{Q4Bio-PhaseEstimation, Q4Bio-HPC, Q4Bio-GuidingState}, as well as the algorithms based on the qubitization framework \cite{Low2019hamiltonian,von2021quantum,goings2022reliably}, we can exploit this example to arrive at general conclusions for the constraints on future quantum computers to achieve a quantum advantage in the general field of atomistic modeling for free energy calculations in biochemistry.

The remainder is organized as follows. In Section \ref{sec:pipeline0} we describe the scientific aspects of the computational pipeline as well as the results from a pipeline run for the anticancer drug-host complex obtained with traditional computational methods.
In Section \ref{sec:promise}, 
we discuss the prospect of future quantum computing methods. Emerging from an analysis of the quantum resources required for impacting the free energy calculation for the ruthenium-based system, we offer general insights for quantum advantage in computational molecular biology.
A conclusion is offered in Section \ref{sec:conclusion}. Additional technical information is provided in the Supporting Information. 

\section{The computational pipeline FreeQuantum}\label{sec:pipeline0}

The partition function $Z$ of a molecular system in the Born-Oppenheimer approximation is given by 
$$  Z=\int d^{3N_{\textrm{nuc}}}R \tr \exp(-\beta H(R)) $$
where $\beta=\frac{1}{k_B T}$ is the inverse temperature with $k_B$ the Boltzmann constant, $T$ the temperature. $H(R)$ is the electronic Hamiltonian for given nuclear coordinates $R$ of the $N_{\textrm{nuc}}$ nuclei. We include in $H(R)$ the electrostatic interaction of the nuclei. At ambient temperature in biological systems higher electronic states can be neglected, resulting in the approximation $\tr \exp(-\beta H(R))\approx \exp(-\beta E_g(R))$ where $E_g (R)$ is the electronic ground state energy
for a fixed nuclear configuration; varying over $R$ gives the potential energy surface (PES).   

Thermodynamic quantities, such as the Gibbs free energy $G$ can be obtained from $Z$ and therefore involve a sampling problem over a vast PES. We do this by traditional molecular dynamics methods, which carve out a discretized trajectory of nuclear positions with help of a force field. 
We can restrict to states with specific properties, e.g., bound or unbound for our protein-ligand system, and indicate this in subscript. 

Our main goal is the computation of the difference
$$  \Delta G_{\text{binding}} = G_{\text{bound}}-G_{\text{unbound}}$$
to high accuracy as this is key to any design attempt that would rely on
differences of $\Delta G_{\text{binding}}$ obtained for different host
molecules (for instance, within a drug discovery campaign).
For each step in the trajectory, we therefore have to compute the ground state energy, i.e., the corresponding point on the PES to high accuracy.
Ideally, we would here employ a full quantum-mechanical energy calculation. As the system consists of thousands of atoms, this is infeasible and we resort to a two-fold quantum embedding strategy resulting in a three-layer quantum-in-quantum-in-classical embedding (QM/QM/MM). 

As in Refs.~\cite{Q4Bio-ML, Q4Bio-Embedding}, we begin by considering the initial force field (MM) which has been used for the trajectory computation. We then choose the
quantum region which will be treated with DFT giving rise to the QM/MM embedding. As the computation of the DFT data is more expensive, we do not compute it for every point on the trajectory but only for some, which we then use to train an ML potential in an active learning loop (ML1). 
To obtain a further refinement, we identify one or more quantum cores within the quantum region giving rise to the QM/QM/MM embedding. The energies of the quantum cores can be computed with traditional quantum chemical methods (currently) or with quantum computing (in the future). In either case, these data points are used to obtain a refined ML potential (ML2) replacing ML1 via transfer learning.

In the following we describe the main aspects of the implementation. First, regarding the free energy perturbation method, second, regarding the ML potentials and, third, regarding the quantum embedding.

\subsection*{Free energy perturbation}

The calculation of the binding free energy $\Delta G_{\text{binding}} = G_{\text{bound}}-G_{\text{unbound}}$
by direct computation of the free energies as outlined above is challenging as the molecular dynamics simulation used in order to sample the phase space is slow and would therefore only explore a very small part. This leads to large uncompensated errors when computing free energy differences, since the sampling for $G_{\text{bound}}$ and $G_{\text{unbound}}$ is over distributions with little overlap.

Free energy perturbation (FEP) 
offers a way of avoiding this difficulty, by telescoping the expression
\begin{align}\label{eq:telescope}    
G_{\text{bound}}-G_{\text{unbound}}=\sum_{k=1}^{s-1} \underbrace{G_{\lambda_k}- G_{\lambda_{k+1}}}_{\Delta G_{\lambda_k}} 
\end{align}
with $s$ steps along a path (e.g., a reaction coordinate) parametrized by $\lambda$ (see e.g., \cite[Fig.~7]{baiardi_quantum_2023}).
Here, $1=\lambda_1> \cdots >\lambda_s=0$, where $0$ and $1$ correspond to the unbound and bound situation \cite[Section 1.3]{lelievre2010free}, and where
$G_{\lambda_k}$ and $G_{\lambda_{k+1}}$ involve very similar sampling when suitable paths are chosen giving rise to error compensation and thus to an effective algorithm (see e.g., \cite{york}). 

An elegant alternative strategy that connects the thermodynamically stable states with a path is to introduce an unphysical parameter into the Hamiltonian and to change this parameter so as to walk along the resulting 'alchemical' path. In protein-ligand binding, this parameter is simply switching off the interaction between protein and ligand: 
$$H_\lambda=H_{\text{protein}}+H_{\text{ligand}}+\lambda H_{\text{interaction}}.$$ 
This defines $G_\lambda$ with 
$G_{\text{bound}}=G_1$ and $G_{\text{unbound}}=G_0=G_{\text{protein}}+G_{\text{ligand}}.$ 

This idea is especially useful as it can be easily adapted to tackle the solvated situation, where it is hard to find a reaction coordinate path due to the presence of the water molecules. Here one first switches off the protein-ligand interaction in the solvated complex. The result corresponds to the sum of the solvated protein and the ligand in vacuum. Considering now the ligand separately, one can switch on the interaction with surrounding water molecules, resulting in
\begin{equation*}
       \Delta G_{\text{solvated binding}} = \underbrace{G_{\text{solvated complex}} - (G_{\text{solvated protein}}+G_{\text{ligand}})}_{\Delta G_{\text{partially solvated binding}}} +\underbrace{(G_{\text{ligand}}-G_{\text{solvated ligand}})}_{-\Delta G_{\text{ligand solvation}}} .
\end{equation*}
The telescoping sum \eqref{eq:telescope}, can be used for each of the two free energy differences on the right, leading to the alchemical FEP method that we implemented \cite{Zwanzig1954, Mey2020, Song2020}. 
Here, the free energy differences $\Delta G_{\lambda_k}$ are determined 
jointly 
using the multistate Bennett acceptance ratio (MBAR) \cite{Shirts2008},
which has a low variance and bias. 
We carefully choose the step sizes as to ensure sufficient overlap of the potential energy distributions along the steps taken on the path, so that the free energies computed with MBAR are reliable.

\subsection*{Machine learning potentials}
We compute three types of free energy differences corresponding to three different, but successively refined, potential energy surfaces. The first corresponds to our standard force field (MM). The second is the ML1 potential for the quantum region combined with the MM potential for the remaining molecular structure. For the third, we replace ML1 by the refined ML2 leading to MM+ML2.

Training a system-specific ML potential from scratch requires a training set with a large number of representative structures. It is common to use both energy and force information (analytical derivatives), and this is done within an active learning loop in order to obtain ML1. 
Starting with 2000 QM/MM data points, as is done in similar ML potentials, we use 90\% of the random structures for training and the rest for validation and train the potential until the root mean square error is converged. We then use this potential for non-equilibrium switching (NEQ), a step in our FEP computation during which new structures
with high uncertainties are encountered. At this point, the active learning kicks
in and new QM/MM reference data for these structures is generated and the training is repeated as described until convergence.

For the second embedding leading to ML2, the active learning is a challenge, as the forces are not directly available in our embedding approaches (see below) and training solely on energies requires significantly more training data as noted in the training of ML1~\cite{Q4Bio-Embedding}. We address this challenge using transfer learning, which allows us to refine ML1 directly with sparse high precision data, a strategy that has previously shown good results \cite{Smith2019, Q4Bio-ML}. To select the structures, we calculate the median energy of the QM/MM energies, and then calculate QM/QM/MM energies for all structures whose QM/MM energies are within 400 kJ/mol
of the median. This is to eliminate artifacts from unphysical structures. 
In total, we used 4570 reference conformers for the protein-ligand complex and 5441 reference conformers for the solvated ligand in order to refine the ML1 potential energy surface in the transfer learning step to ML2.

\subsection*{Quantum embedding}
In the following we will explain the construction and choices made in the quantum-mechanically refined PESs. To begin with, the free energy sampling is conducted via molecular dynamics simulations on the PES corresponding to a standard force field with the equilibrated structure as a starting point. The energies and forces are expected to be inaccurate in certain areas of the PES due to quantum-mechanical effects. 
It is therefore desirable to switch to a quantum-mechanical description of the protein-ligand system. Since the system is too large, this is out of reach, even in principle, with traditional, high-accuracy classical computation. Such systems could, in principle, be represented on a quantum computer, but they are expected to remain out of reach in practice for at least the coming decades.
A full quantum-mechanical representation of the system, however, is also not needed, as the relevant quantum-mechanical effects in biochemical systems are localized. Furthermore, the errors in some localized regions are likely to be the same throughout all of the calculations, therefore leading to error cancellation since we are looking at free energy difference. In the case of protein-ligand binding, the most relevant region is the interface of protein and ligand, in our case including the ligand as it is small. 

Quantum embedding methods allow us to define and incorporate a quantum region within a larger classical environment. The total energy is then the sum of the energy of the outer part ($E_{\text{MM}}$) and the ground state energy of an electronic structure Hamiltonian of the quantum region in an effective MM potential from the outer part 
\begin{align}
\label{eq:embedding}H_{\text{quantum region}}=H_{\text{electronic, quantum region}}+V_{\text{interaction, MM}}.
\end{align}
As the required quantum region would still involve thousands of orbitals, an exact quantum-mechanical treatment remains out of the question for traditional methods and also for the foreseeable future on quantum computers. An approximate treatment with DFT is possible but leaves us with large and uncontrolled errors which are not systematically improvable. As those DFT errors are expected to be further localized, we resort to an additional quantum-in-quantum (QM/QM) embedding. Here, we identify one or more smaller 'quantum cores' inside the quantum region which we treat with accurate wavefunction methods embedded in the larger quantum region which we treat with DFT. The Hamiltonian for the core consists of the electronic part plus an additional external potential coming from the quantum-in-quantum embedding:
$$H_{\text{quantum core}}=H_{\text{electronic, quantum core}}+V_{\text{interaction, DFT}}.$$
Thus we obtain a quantum-in-quantum-in-classical embedding (QM/QM/MM). 

Whereas it is desirable to let the quantum region cover the entire protein-ligand interface which includes the small ligand, we here focused the quantum region on the ligand only in our ruthenium-compound pipeline run. We note that this still incorporates some of the quantum-mechanical interaction effects between host and ligand due to the external potential on the ligand (see \eqref{eq:embedding}). We have also chosen the quantum region and the quantum cores manually, but note that an automated algorithm for the quantum region selection is implemented and can be used in the future \cite{Brunken2021, Q4Bio-Embedding-3b}.

\subsection*{Quantum engines}
\label{sec:quantum-engines}
The pipeline is set up such that the size of the quantum cores, to be treated with wavefunction-based methods, can be adjusted both to the protein-ligand complex at hand, but also to the computational resources available. At present the computational resources consist of high-performance computers with which we have carried out quantum chemical calculations as we discuss below. The curse of dimensionality will ultimately limit the ability of these computations to capture the correlations of large quantum cores, and at this point it would be advantageous to replace them with quantum computations. The development of both quantum algorithms and quantum hardware is rapid and has the potential to be a game changer within our approach as we detail below.

In the full run of our pipeline on the ruthenium drug-protein complex, we have used a Huzinaga-type projection embedding  \cite{Q4Bio-Embedding}. Here, a single core is created inside the quantum region. Since not all of its orbitals can be considered in a single energy calculation due to its large size, we have resorted to a complete active space (CAS) approach, which can deliver the static correlation with help of a full configuration interaction (FCI) wavefunction in a restricted space of orbitals \cite{Helgaker2013}. The choice of active orbitals has been done with the fully automated autoCAS algorithm \cite{Stein2016}. We rely on the density matrix renormalization group (DMRG) as a proxi for the full configuration interaction (FCI) calculation \cite{Baiardi2020}. To account for the dynamic correlations that cannot be captured within the CAS realm, we use second-order N-electron valence state perturbation theory (NEVPT2) \cite{Baiardi2020}.

To achieve high flexibility for the application of quantum algorithms, FreeQuantum can employ a second, complementary, embedding strategy based on bootstrap embedding \cite{Liu2023Bootstrap,Q4Bio-Embedding-3b}. Here, several smaller overlapping quantum subsystems are chosen within one quantum core, each of which is subjected to an energy calculation before being embedded in a consistent way into the (larger) quantum core. The size of these subsystems, however, affects the accuracy (see the Supporting Information Fig.~\ref{fig:be_corr_plot} for results on our Ru-drug example), but this can, in fact, be increased with future quantum computing methods.

In general, both the Huzinaga-based projection approach and the bootstrap approach can exploit quantum computational resources in different ways. For a multi-configurational (strong correlation) problem, for instance, projection embedding would allow FreeQuantum to choose an active orbital space within a quantum core for an exact diagonalization approach with subsequent treatment of the remaining dynamic correlation, whereas bootstap embedding would dissect the quantum core into smaller overlapping subsystems, so that all orbitals per subsystem can be subjected to exact diagonalization (rendering a dynamic-correlation treatment unnecessary).

\subsection*{Ruthenium-based anticancer drug}
\begin{figure}[t]
  \centering
    \begin{subfigure}[b]{0.33\textwidth}
    \vspace{0pt} 
    \centering
    \includegraphics[width=\textwidth]{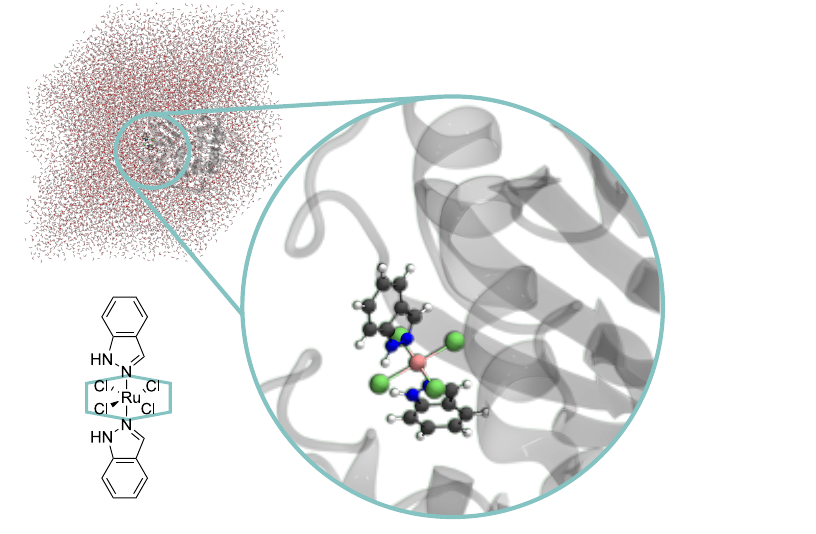}
    \caption{}
    \label{subfig:guest-host}
  \end{subfigure}
  \hfill
    \begin{subfigure}[b]{0.63\textwidth}
    \vspace{0pt} 
    \centering

  \begin{threeparttable}
\footnotesize
    \begin{tabular}{lll}
      \toprule
      Method & MCL-1/19G & GRP78/NKP-1339 \\
      \midrule
      MM  & $-37.5\pm0.4$ \cite{Q4Bio-ML} & $-19.1\pm1.5 $ \cite{Q4Bio-ML} \\
      PBE/MM  & $-35.3\pm1.8 $ \cite{Q4Bio-ML} & $-17.0\pm2.6$ \cite{Q4Bio-ML} \\
      UMP2/PBE/MM  & \qquad \ \ ---   & \, $-9.6\pm2.2$ \ this work  \\
      LCCSD(T)/PBE/MM              & $-37.2\pm1.0$ \cite{Q4Bio-Embedding} & \qquad \ \ ---                   \\
      UCCSD(T)/PBE/MM              &  \qquad \ \ ---                      & $-10.8\pm2.4$  \ this work  \\
      NEVPT2/PBE/MM & \qquad \ \ --- &  $-11.3 \pm 2.9$  \ this work \\
      Experiment                          & $-37.3\pm0.1$ \cite{Friberg2012} & \qquad \ \ ---   \\
      \bottomrule
    \end{tabular}
     \end{threeparttable}
\caption{}
\label{subfig:results}
  \end{subfigure}

    \caption{(a)
        Host-guest complex of the chaperone BiP (Binding immunoglobulin Protein) (the protein host) with the ruthenium transition-metal complex (the small-molecule guest, highlighted in the blue circle in ball-stick representation with carbon atoms in grey, chlorine atoms green, nitrogen atoms blue, hydrogen atoms white, and Ru in orange; Lewis structure given in the lower left corner). The QM region is in the QM/MM model is the complete Ru drug molecule, whereas the
        quantum core for the QM-in-QM embedding is highlighted in the blue hexagon of the Lewis structure.
        (b) Overview of the binding free energies (in kJ/mol) obtained at different stages of the FreeQuantum pipeline and with different quantum chemical methods to treat the quantum core (see Methods, also for an explanation of all acronyms). 
        Next to the results for the ruthenium drug-protein complex we provide our previous results for 
        another protein-ligand system, the myeloid cell leukemia 1 (MCL-1) protein, dysregulation of which is associated with various cancers, inhibited by the small organic closed-shell molecule 19G for which classical force fields (MM) work very well 
        and an experimental reference is available. The error bars are determined as in Ref.~\cite[eq.(14)]{Q4Bio-Embedding}.}
    \label{fig:embedding}
\end{figure}

We have run the full FreeQuantum pipeline on the binding process of a small-molecule drug to a protein. The chosen molecule NKP-1339 (synonyms are IT-139, KP-1339) \cite{Trondl2014, Flocke2016} causes cell death through apoptosis via several mechanisms, one of which is the inhibition of the Hsp70-family molecular chaperone BiP (Binding immunoglobulin Protein), which regulates the unfolded protein response \cite{Bakewell2018,bertolotti2000dynamic}. See Fig.~\ref{subfig:guest-host} for an illustration. 

As a case of molecular recognition of a small molecule by a large biomacromolecule, which always requires sampling of the configuration space accessible to the emerging host-guest complex and comparing it to the configuration space relevant for the separated molecules in solution, the chosen case represents, as desired, the high-configurational situation. 
Metal atoms and their coordination sphere are usually hard to describe with classical force fields. The ruthenium transition metal complex, furthermore, makes the whole system open shell. The electronic ground state is a spin doublet, further highlighting the need for a quantum-mechanical description. The resulting metal-drug protein complex, which belongs to the top-left in the biomolecular quantum simulation quadrangle, is therefore an excellent example to demonstrate the developed FreeQuantum pipeline.

The full run predicts binding with a binding free energy of 
$\Delta G_{\mathrm{binding}}^{\mathrm{MM}} = -19.1 \pm 1.5$\,kJ/mol for a pure MM description and
$\Delta G_{\mathrm{binding}}^{\mathrm{MM+ML1}} = -17.0 \pm 2.6$\,kJ/mol for the first embedding approach.
The QM description in this first embedding hybrid model was based on DFT with all shortcomings of approximate exchange--correlation density functionals. Therefore, it is an excellent case for the three-layer QM/QM/MM description, for which we obtained $\Delta G_{\mathrm{binding}}^{\mathrm{MM+ML2}} = -11.3 \pm 2.9$\,kJ/mol for our most advanced wavefunction methods for the core (NEVPT2 on top of CAS configuration interaction, CAS-CI). 
Fig.~\ref{subfig:results} presents those results together with the data from a QM/MM embedding based on DFT \cite{Q4Bio-ML} (right column). 
Fig.~\ref{fig:energy-distributions} in the Supporting Information presents the distribution of energies of the embedded cores for first and second embedding as well as their difference, illustrating that the embedding strategy fundamentally changes the energy landscape.   

In Fig.~\ref{subfig:results} we also provide in the middle column corresponding data for a different system involving an organic closed-shell ligand denoted MCL-1/19G \cite{Q4Bio-ML,Q4Bio-Embedding}. We have used this system, for which reliable classical force field and experimental results are available, in earlier work as a benchmark to demonstrate that our nested embedding approach based on accurate coupled cluster data reproduces this reference. For the ruthenium drug we here find a coupled cluster result that deviates by about 6 to 8 kJ/mol from the DFT and MM results, respectively. At the same time, the NEVPT2 result agrees with the coupled cluster data (-10.8 kJ/mol for unrestricted coupled cluster and -11.3 kJ/mol for NEVPT2), indicating that both electron correlation methods achieve the same high accuracy for this system. Hence,
with our best first-principles approaches, we obtain a free energy of binding for the ruthenium drug of about -11 kJ/mol, which we may consider as a prediction to be challenged by experimental work in the future.

 Although our strategy allows us to take into account and demonstrate the impact of high accuracy quantum chemical calculation in a biologically relevant case, it also highlights its limitations due to the curse of dimensionality. 
 First, any wavefunction-based traditional method lacks guarantees on the energies which may therefore compromise the accuracy of the final free energy. Second, it will be severely limited in its applicable size. As we will emphasize, quantum computation has the potential to overcome both problems and thus lead to biological quantum advantage. We will provide estimates of the size, quality and speed of future quantum computers to realize this potential.

\section{The potential of quantum computing}
\label{sec:promise}

In the future, the energy computation of the quantum cores may be carried out with quantum algorithms on quantum computers. Indeed, one of the main motivations of the FreeQuantum pipeline is to anticipate the potential of quantum computation for electronic structure computation and the application to biological problems. In this section, we discuss the options and prospects and give concrete resource estimates for the ruthenium drug-protein complex.

The number of high-accuracy energy data points required depends on uncertainty quantification during the machine learning. For the chosen complex this is found to be around 4000. The required energy accuracy is chemical accuracy (i.e., about 1 kJ/mol), but generally depends on the region where the sampling happens and is again determined by the uncertainty quantification of the ML potential.

\begin{figure}[!ht]
  \centering
  \begin{subfigure}[b]{0.4\textwidth}
    \vspace{0pt} 
    \centering
    \includegraphics[width=\linewidth]{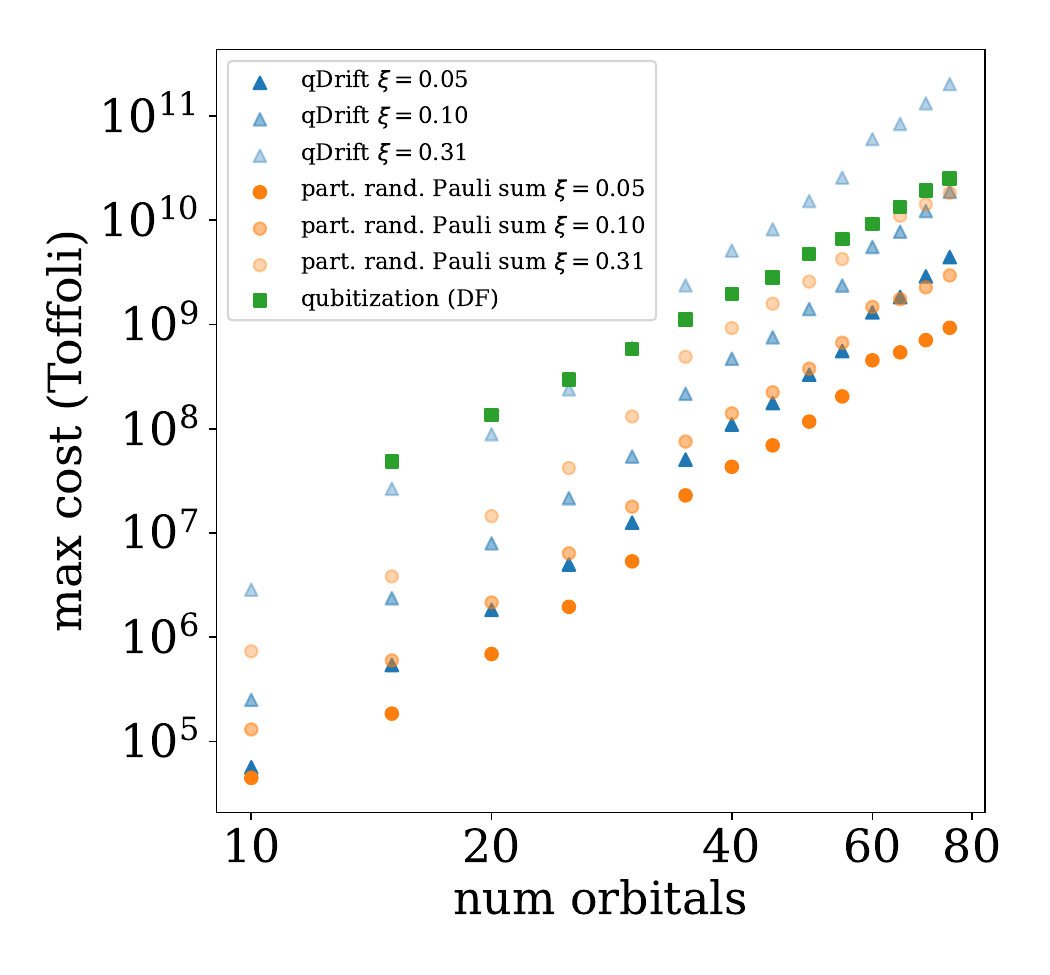}      
    \caption{}
    \label{}
  \end{subfigure}
  \begin{subfigure}[b]{0.4\textwidth}
    \vspace{0pt} 
    \centering
\includegraphics[width=\linewidth]{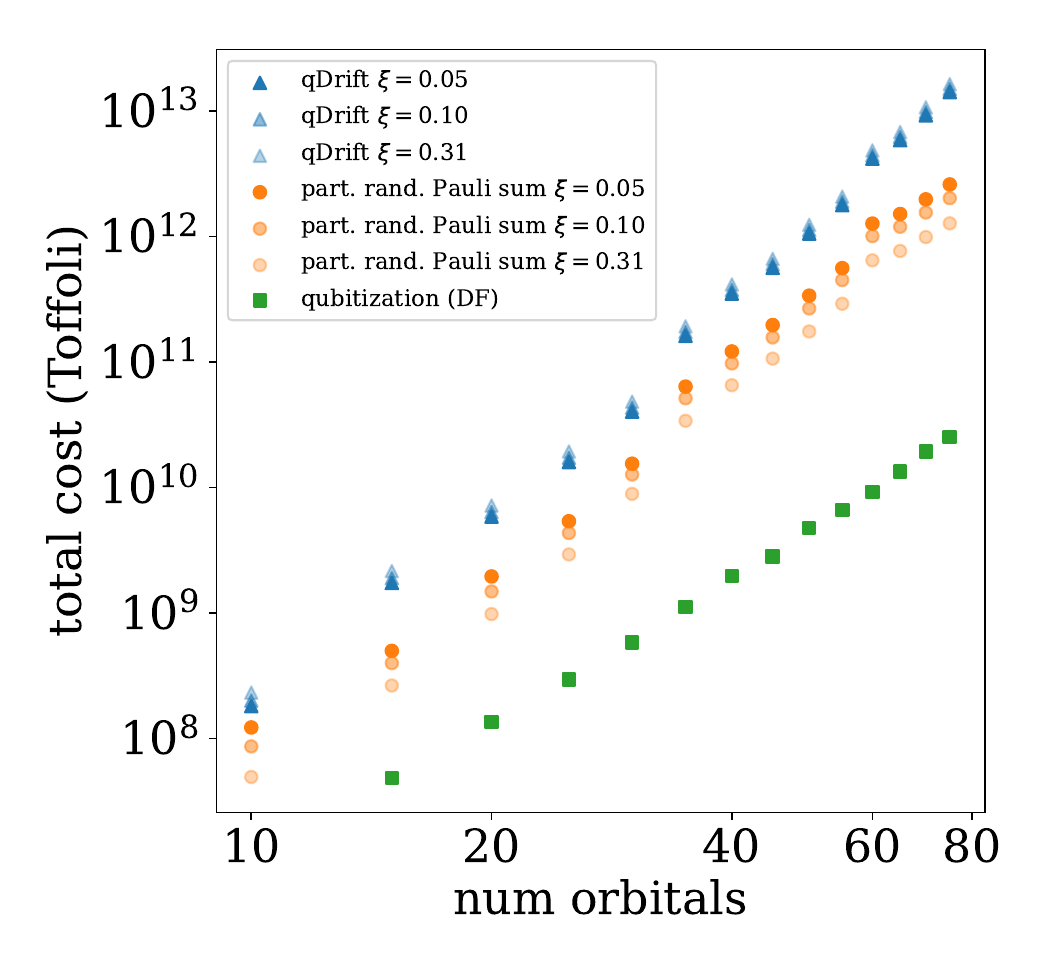}  
\caption{}
  \end{subfigure}
  \\[1em]
      \begin{subfigure}[b]{0.4\textwidth}
    \vspace{0pt} 
    \centering
    \includegraphics[width=\linewidth]{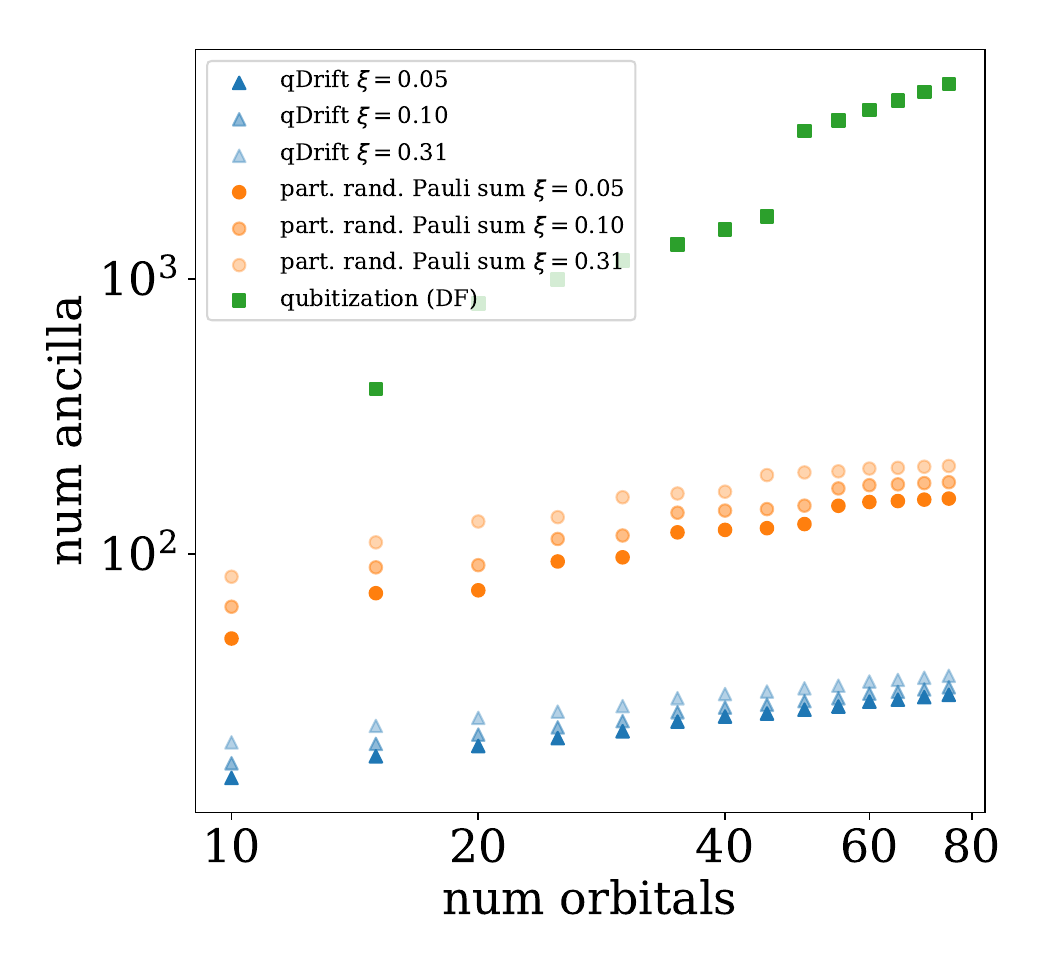}      
    \caption{}
    \label{}
  \end{subfigure}
  \begin{subfigure}[b]{0.4\textwidth}
    \vspace{0pt} 
    \centering
\includegraphics[width=\linewidth]{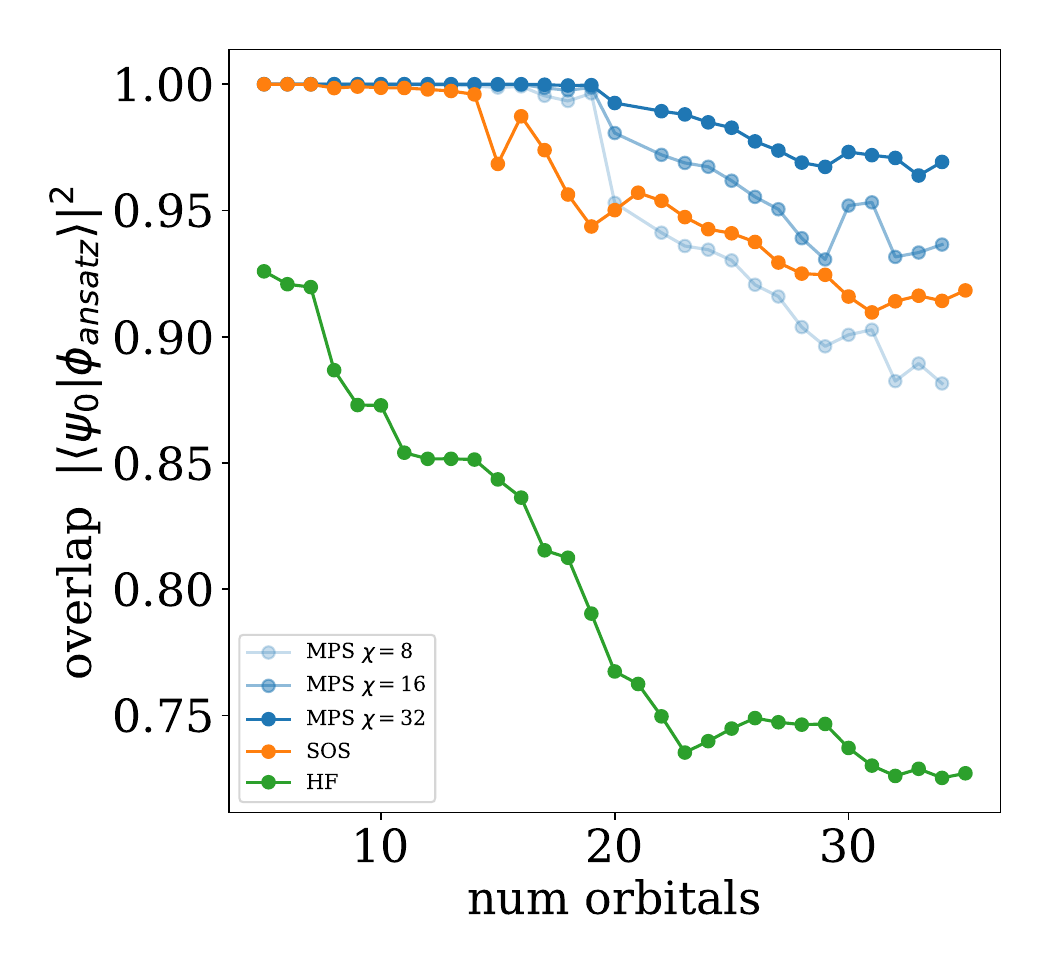}  
\caption{}
  \end{subfigure}
  \caption{(a)  Maximal number of gates per circuit with target accuracy $0.001$ Hartree for active spaces of the Ruthenium ligand for partially randomized Trotter (in dependence of overlap) and qubitization. $\xi=\arcsin{\frac{1-\eta}{\eta}}$, where $\eta$ is the overlap. (b) Total number of gates per circut with target accuracy $0.0016$ Hartree for active spaces of the Ruthenium ligand for partially randomized Trotter (in dependence of overlap) and qubitization. 
  (c) Number of qubits needed in dependence of the above algorithms and overlaps. The product-formula based methods can avoid all ancilla qubits, at an increased cost of synthesizing single-qubit rotations.
  (d) Overlap with ground state for different types of guiding states for the Ruthenium system with increasing number of orbitals in the active space. As proxy for the ground state we used a converged DMRG state. In orange the Sum-of-Slater (SOS) overlap with $4 \times N$ Slater determinants where $N$ is the number of orbitals of the active space. In blue the overlap of a matrix-product state (MPS) obtained by truncating the DMRG state to the indicated bond dimension $\chi$, and in green the overlap with the Hartree-Fock state (HF). This indicates that good overlap states that can be prepared on a quantum computer exist for our system of interest.
 (a-d) For details regarding the graphs, see Supporting Information.}
    \label{fig:ruth-depth}
    \label{fig:overlap-ruthenium}
\end{figure}

The algorithmic framework for computing the ground state energy of the quantum core Hamiltonian is quantum phase estimation (QPE).
While there exist variational alternatives for ground state energy computation such as the variational quantum eigensolver (VQE), for generating training data the accuracy guarantees of QPE are valuable, and VQE suffers from fundamental computational limitations for larger systems \cite{larocca2024review}.
QPE requires the preparation of a guiding state $\ket{\psi}$ on the quantum computer with significant ground state overlap, and simulation of the Hamiltonian evolution on the quantum computer. For Hamiltonians in biochemistry, finding guiding states with high ground state overlaps is possible in many cases \cite{Q4Bio-GuidingState,fomichev2023initial,berry2025rapid} and we also confirm this for the described choice of quantum cores for our ruthenium system.
This is shown in Fig.~\ref{fig:overlap-ruthenium}.
We find that the Hartree-Fock state (which is trivial to prepare on a quantum computer) provides sufficient overlap even on fairly large active spaces. Higher overlap is obtained by considering guiding states consisting of a linear combination of a small number of Slater determinants. Low bond-dimension matrix product states (MPS), which can easily be prepared \cite{fomichev2023initial,berry2025rapid}, 
achieve very high overlap.

The main computational cost derives from the Hamiltonian simulation, which depends on the required precision and the number of orbitals.
We consider two regimes, the first is a regime in which the number of qubits limited (note that $2N$ qubits for are required to represent a state on an active space of $N$ spatial orbitals) and a second regime, where a significant number of qubits can be used as ancilla qubits. In the first regime, we consider single-ancilla QPE in conjunction with partially randomized product formula (Trotter) methods \cite{Q4Bio-PhaseEstimation} and qDRIFT \cite{Campbell2019Random, Q4Bio-PhaseEstimation}.
If there is a high overlap guiding state, this can be used to reduce the number of gates per circuit (at the expense of running more circuits), allowing for some degree of parallelization and milder noise requirements, 
which may be valuable for early fault-tolerant hardware. In this example, such states are indeed available, making this a relevant trade-off (but we note that ideally also the classical computational cost of finding these states should be taken into account).
In Fig.~\ref{fig:ruth-depth} we show the maximal number of gates per circuit, as well as the total cost, needed for the ruthenium system with varying active space size.
These estimates require knowledge of the Trotter error and are informed by a calculation of the Trotter error on this system carried out on HPC and GPU clusters with exact state vector simulations of up to 36 qubits~\cite{Q4Bio-HPC}. In the second regime, when a significant amount can be used as ancilla qubits, qubitization methods outperform product formula based methods when total runtime is considered; resource estimates for qubitization are also shown in Fig.~\ref{fig:ruth-depth}.

The runtime resulting from the gate counts reported in Fig.~\ref{fig:ruth-depth} depends strongly on the hardware as well as the fault-tolerant architecture.
In order to get a rough idea of the concrete quantum resources required, let us see what is required in order for a single energy computation to be performed within $20 \, \mathrm{min}$ (which would correspond to two months for 4000 data points of training data of the pipeline).
We first consider the case of an active space of 30 spatial orbitals, which has previously been identified as a point where comparison to traditional methods becomes relevant \cite{goings2022reliably}.
In this case, the Trotter-based algorithm requires at least 60 qubits to represent the state and gate errors below $10^{-7}$, see Fig.~\ref{fig:ruth-depth}. A computation time of $20 \, \mathrm{min}$ per energy can be reached if the average gate time is below $10^{-7}\, \mathrm{s}$. Outperforming DMRG will require larger active spaces; here we consider 60 spatial orbitals. While product formulas have benefits in terms of the number of logical qubits and the possibility of breaking up into smaller circuits, at this system size the total runtime becomes the dominating factor and qubitization-based methods are more relevant.
With around 1000 logical qubits, gate errors below $10^{-10}$ and an average gate time of $10^{-7} \, \mathrm{s}$, the qubitization resource estimates reach the $20 \, \mathrm{min}$ target.
Further optimization of qubitization-based methods is likely to reduce this cost by one to two orders of magnitude \cite{caesura2025faster,low2025fast}.
In order to fully take into account dynamical correlation, we could either increase the number of orbitals further, thereby avoiding NEVPT2 altogether, or, use quantum algorithms to compute such perturbative corrections \cite{Q4Bio-MoreQuantumChemistry}. 

To put these numbers in perspective at the hardware level, current parameters for superconducting qubits are on the order of $5\times 10^{-8} \, \mathrm{s}$ gate time and $5 \times 10^{-3}$ error \cite{Kjaergaard2020}. In ion-based platforms, two-qubit gate times of around $10^{-4} \, \mathrm{s}$ at gate error $3 \times 10^{-4}$ have been reported \cite{ionics}. Using fault-tolerant techniques to achieve the required accuracy will lead to an increased average time per logical gate, showing that parallelization and low time-overhead fault-tolerant operation will be crucial for sufficiently fast operation.

 A parallel computation of the energies is possible in our current implemented pipeline if several quantum computers are available, allowing to reduce the total wall-clock runtime. 
 We note that to compare favorably to experimental efforts, it would be required to execute the entire pipeline within a time frame of $24 \, \mathrm{h}$, which could reasonably be achieved by parallelization.
 If active learning is used in the ML2 training, parallelization of the energy data may be limited.
 Additionally, as detailed above, for each energy and sufficiently large ground state overlap one can parallelize the estimate (especially when using randomized product formulas).

\section{Conclusions}
\label{sec:conclusion}

An accurate quantum-mechanically informed assessment of the interactions between molecules in molecular recognition processes can deliver a sound basis for (i)~its detailed physics-grounded understanding, (ii)~an accurate estimate of the binding free energy, and (iii)~a flexibility that guarantees transferability of a methodology to arbitrary molecule classes. However, 
a complete computational pipeline capable of routine application has so far been missing. 
In this work, we have addressed the feasibility challenges in a software pipeline by
combination of various innovations: 
(i)~A two-fold embedding strategy allows us to carve out quantum cores of variable size from a large quantum region so that highly accurate quantum calculations can be carried out for them in order to obtain high-accuracy data for those regions.
(ii)~These quantum calculations can be performed with traditional methodology (e.g., based on coupled cluster or multi-configurational approaches) or with future fault-tolerant quantum computers.
(iii)~To be able to deal with scarce high-accuracy data, we exploit an efficient transfer machine learning approach that threads the quantum data into the first level of embedding (that is, into the quantum-classical hybrid model).
(iv)~An element-agnostic machine learning strategy allows us to sample the free energy of binding at all levels 
(quantum-classical and quantum-in-quantum embedded in classical) of embedding.
(v)~For the innermost quantum-in-quantum embedding, we investigated two different embedding strategies to highlight the modularity and flexibility of the overall approach. The first embedding scheme is based on Huzinaga-type projection embedding, the other one on bootstrap embedding. Hence, a high degree of versatility is guaranteed for the connection of the structures from classical configuration space sampling to the accurate quantum calculations for the quantum cores of individual configurations.

The whole pipeline enables us to tailor the quantum core size (measured in terms of atoms and/or orbitals) to the capabilities of the available accurate quantum engines.
Moreover, our pipeline exhibits a high degree of autonomy and it is highly modular (i.e., different modules for specified tasks can be easily exchanged) so that future improvements of the technology can be easily incorporated. When the required quantum computational resources as outlined become available, they can directly be applied within the FreeQuantum pipeline. We have done concrete resource estimates for the 
ruthenium drug-protein complex, which is located in the 
top-left of the biomolecular simulation quadrangle.
We had chosen this complex, as standard force fields struggle because it is an open-shell spin doublet, but it was still meaningful to demonstrate the pipeline with traditional HPC calculations. It would have been impossible to attempt a case of the top-right of the quadrangle. Note, however, that the resource estimates for quantum phase estimation are dominated by the Hamiltonian simulation, which is determined by the size of the problem and not by the ground state correlation structure. Although finding and preparing good guiding states will become more challenging (as they involve traditional quantum chemical methods), we expect our resource estimates for the ruthenium drug-protein complex 
to be of similar order of magnitude for cases in the top-right of the quadrangle. 

From previous work on quantum algorithms for electronic structure computations in realistic cases one might have concluded that applying quantum computers in free energy computations is unrealistic for the foreseeable future due to the extensive configurational sampling. By reducing the size of the quantum region through a multilayer quantum embedding strategy paired with efficient use of quantum data through machine learning, we change this perspective. We conclude that there is a realistic prospect of obtaining useful results not only in strongly correlated regimes, but also where the correlations are not so strong, the vast majority of biological cases. Furthermore, though our work dealt with molecular recognition, the methodology is general and can be used for any atomistic simulation problem in biochemistry replacing less accurate classical force fields by the ML1 and ML2 potentials in our hybrid model approach.

Our main contribution with this work is therefore to provide a realistic and concrete way to use quantum computers for the computation of free energy computations and thereby open a research direction leading to future quantum advantage in biology. When this exactly happens in terms of more precise estimates of free energies  will depend on the break-even point with 
high-accuracy traditional electronic structure approaches in case of strongly correlated systems or other wavefunction based methods (e.g., coupled cluster) in case of weaker correlations. What is clear, however, is that quantum computing has the potential to change the foundation of free energy calculations by delivering certified-quality useful training data within the pipeline.
FreeQuantum is open source and free of charge, welcoming both development contributions as well as deployment for the benefit of biochemistry and the pharmaceutical sciences.

\section*{Acknowledgments}
We acknowledge funding from the Research Project ``Molecular Recognition from Quantum Computing''. Work on "Molecular Recognition from Quantum Computing" was supported by Wellcome Leap as part of the Quantum for Bio (Q4Bio) Program. Furthermore, M.\ C., J.\ G., T.\ W., F.\ W., M.\ B., M.\ E., R.\ T.\ H., M.\ M., M.\ R. V.\ S., M.\ E. and A.\ H. acknowledge support from the Novo Nordisk Foundation (GrantNo. NNF20OC0059939 `Quantum for Life'). M.M. acknowledges support from the novoSTAR Programme by NovoNordisk A/S. G.\ C.\ S. and W.\ B.\-J. acknowledge support from the European Research Council (ERC) under the European Union's Horizon 2020 research and innovation program (grant agreement No 865870). G.\ C.\ S. and M.\ S.\ T.\ acknowledge support from the Novo Nordisk Foundation, grant number NNF22SA0081175, NNF Quantum Computing Programme and NNF20OC0060019, SolidQ. A.\ K. acknowledges support from the Novo Nordisk Foundation NNF20OC0062606 and NNF20OC0063268. T.\ W., R.\ T.\ H. and M.\ R. acknowledge support from the NCCR Catalysis (grant number 180544), a National Centre of Competence in Research funded by the Swiss National Science Foundation. M.\ R. also acknowledges financial support from the Swiss National Science Foundation through Grant No. 200021\_219616. M.\ E. acknowledges an ETH Zurich Postdoctoral Fellowship. The work is co-funded by the European Union (ERC, DynaPLIX, SyG-2022 101071843, to K.L.-L.). Views and opinions expressed are however those of the authors only and do not necessarily reflect those of the European Union or the European Research Council. Neither the European Union nor the granting authority can be held responsible for them.

\newpage 
\appendix

\section*{Supporting Information}\label{sec:methods}

In the following we provide further information on the workflow of the FreeQuantum computational pipeline, on computational aspects thereof, as well as the quantum computational resource estimates.

\subsection*{Workflow of FreeQuantum}
\label{sec:pipeline}

\begin{figure}[h]
    \centering
    \includegraphics[width=0.7\linewidth]{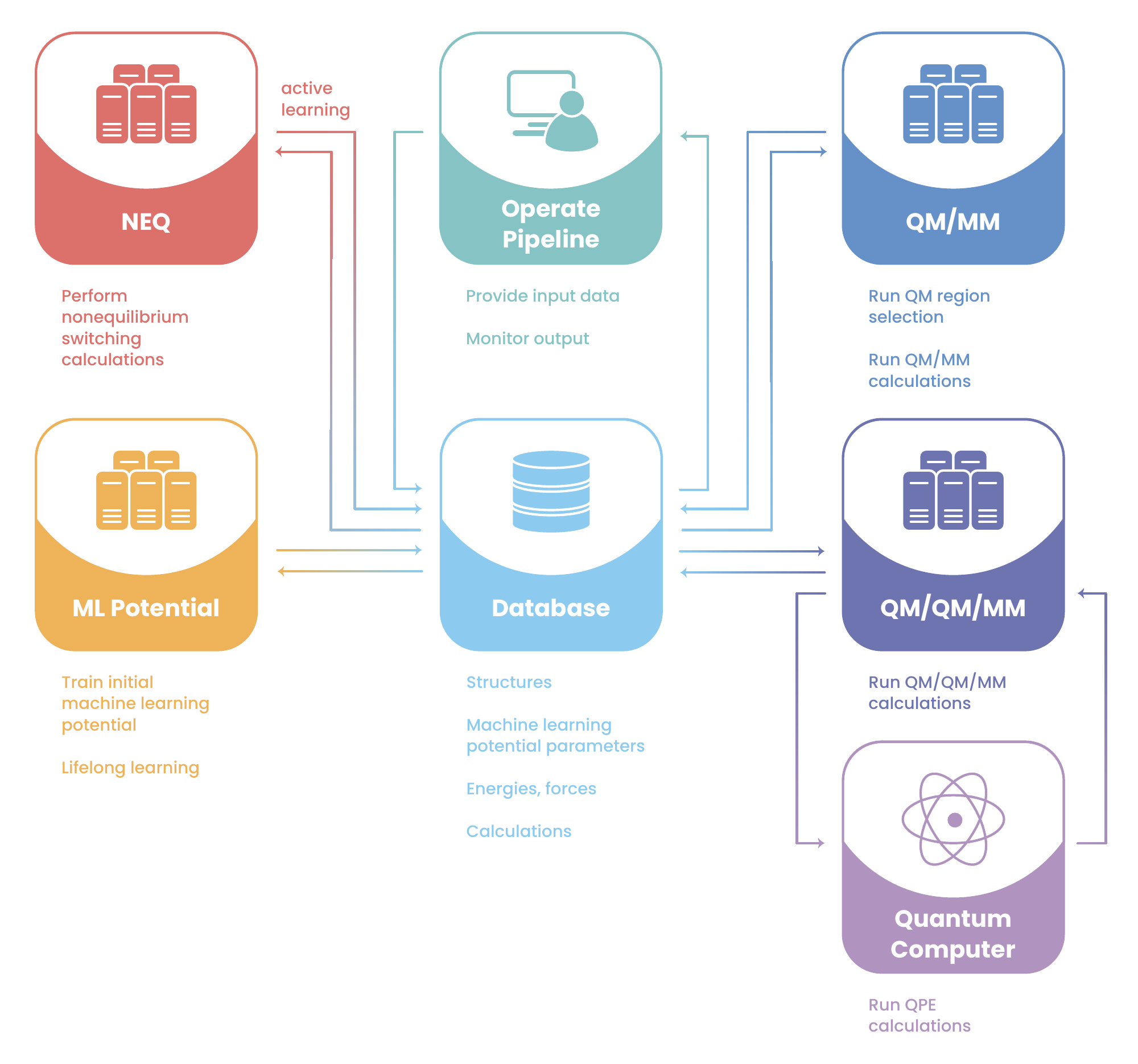}
    \caption{\label{fig:data_flow} \small Data flow in the FreeQuantum pipeline. The database (depicted in the center) facilitates the exchange of data between the individual modules of the pipeline,
        which are shown on the left- and right-hand sides.
        The QM/MM and QM/QM/MM modules enable the calculation of
        accurate reference data.
        The ML potential module
        creates an ML potential from these reference data, which is then used by the NEQ module to calculate the binding free energy.
        The pipeline is
        controlled by a human operator (bottom center).}
\end{figure}

A schematic representation of the data flow within the FreeQuantum pipeline is shown in Fig.~\ref{fig:data_flow}. 
All steps requiring significant human time are fully automated (with the exception of the initial structure preparation step). However, the individual steps can also be triggered manually via simple Python scripts. Such a manual operation can allow for a close monitoring
of the individual steps.

All relevant data are stored in a central MongoDB database \cite{mongodb}. The software modules which
collectively form the pipeline communicate through this database. This architecture allows for a
geographically distributed deployment of the individual pipeline modules: literally, any module can run on
any computer in the world, as long as it is able to communicate with the database.

After the first, purely MM-based FEP calculation, snapshots (typically on the order of 2,000) from the resulting trajectory are extracted via one of the pipeline operation scripts and stored in the database. Optional calculations to automatically determine the QM region based on a
few (typically about 100) of these snapshots can be set up if no intuitive QM region selection is available. These calculations
are then also stored in the database (i.e., all necessary input to carry out these calculations is stored). Then, these calculations are carried out by the QM/MM calculations module. Once the QM region has been determined, QM/MM calculations are set up for all snapshots stored in the
database, and subsequently carried out by the QM/MM calculations module. The resulting energies and forces are stored in the database.

These energies and forces are then read from the database by the ML potential module, and used as input data to train an ML potential (ML1). The resulting parameters are written back to the database so that a NEQ calculation to refine the potential energy surface with the ML force field can be carried out (this is done by the NEQ calculations module). During the molecular dynamics simulation necessary to carry out the NEQ calculation, structures might be encountered for which the ML force field indicates a large uncertainty. If this should be the case, these structures are reported so that additional QM/MM calculations can be carried out to provide the necessary training data to improve the ML1 potential.  

The QM/QM/MM module works along the same concept as the QM/MM module. It
retrieves calculations, which are to be carried out, from the database and writes back the resulting energies and, in principle,
forces (which will be used to train the ML2 potential). We note that here, for the training of ML2, the presented calculations did not use forces, but could
easily be adapted to do so, if the corresponding high accuracy quantum data become available. 

Moreover, we note that the QM/QM/MM module can also
interface to a quantum computer engine in order to carry out QPE calculations on the quantum cores. This
is because, naturally, a quantum computer cannot directly communicate with the database, but instead
has to rely on a classical computer for this. Instead of creating a separate communications module, we
opted to integrate this within the QM/QM/MM module. Besides simplifying the overall
architecture, this design has the advantage that the input needed to carry out a QPE calculation does not
have to be stored in the database. It is worth pointing out that a second-quantized Hamiltonian needs all
corresponding one- and two-electron integrals to be fully defined; these integrals can quickly
demand a lot of storage space.

Due to the concept of communicating exclusively through a central database, the only requirement that a
module must fulfill to be compatible with our pipeline is that it must be able to communicate with
the database. We provide a wrapper \cite{database130} to simplify this communication. Therefore, one
has a lot of freedom when building a module, enabling one to construct it exactly according to the
specific needs of this module. For example, the QM/MM and the QM/QM/MM modules need a
large software stack to carry out their calculations, which can significantly complicate deployment.
Therefore, we realized these modules as Apptainer (formerly Singularity) images which are
straightforward to share among research groups and to launch on high-performance computing environments.
The actual workflow to carry out QM/MM and QM/QM/MM calculations is implemented in SCINE Puffin \cite{puffin130},
which is a framework to carry out complex computational tasks, as recently described elsewhere \cite{Weymuth2024}. The whole source code of the FreeQuantum pipeline will be made available free of charge and open source on github upon publication of this work.

\subsection*{Computational aspects of FreeQuantum}

The model construction for the Ru-based host-guest complex alongside initial molecular mechanics and quantum-classical hybrid data and the foundational machine learning approach have been reported in Ref. \cite{Q4Bio-ML}. Ref. \cite{Q4Bio-Embedding} described the principles of our transfer learning approach for the Huzinaga-type quantum-in-quantum embedding, albeit for a different protein-guest complex that is not as challenging
as the Ru-based one considered in this work.

The molecular structures of the protein-ligand complex of the Ru-based anticancer compound NKP-1339 \cite{Trondl2014, Peti1999} bonded to the heat shock protein GRP78 \cite{Macias2011} (PDB entry 3LDO) were taken from the MM FEP endpoint trajectories and the active learning structures in Ref.~\cite{Q4Bio-ML}. For each structure, we calculated the electronic energy with Huzinaga-type embedding based on the ansatz described in Ref.~\cite{Q4Bio-Embedding}. Huzinaga-type embedding \cite{Hegely2016} is a variant of projection-based embedding \cite{Manby2012} allowing the combination of nearly arbitrary electronic structure methods by partitioning the occupied orbital space. Huzinaga-type embedding is formally exact within the framework of Kohn--Sham DFT, making the approach robust even if the system partitioning cuts covalent bonds.

In our Huzinaga-type multilevel approach, we combined unrestricted second order M{\o}ller-Plesset perturbation theory (UMP2), the strongly contracted NEVPT2 \cite{Angeli2001, Angeli2002}, or unrestricted CCSD(T) (UCCSD(T)) with Kohn--Sham DFT using Perdew, Burke, and Ernzerhof's exchange--correlation functional PBE \cite{Perdew96} with Grimme's D3 dispersion correction \cite{Grimme2010a} and Becke--Johnson damping \cite{Grimme2011} using the def2-SVP basis set \cite{Ahlrich2005} and the corresponding effective core potential for Ru \cite{Andrae1990}. The MM environment was parameterized using the Amber ff99sb*-ILDN force field \cite{best2009optimized,Lindorff‐Larsen2010}.

The MP2, NEVPT2, and unrestricted CCSD(T) calculations were performed with the program PySCF \cite{Sun2018}, taking the valence orbital coefficients and one-particle Hamiltonian from Serenity \cite{Serenity2018, Niemeyer2022} as input. The actual embedding was performed with Serenity. In addition to the orbital partitioning from Ref.~\cite{Q4Bio-Embedding}, we relaxed the orbitals in the quantum core in an embedded self-consistent field calculation. Relaxing and canonicalizing the orbitals was required by the UMP2 and NEVPT2 implementations in PySCF. 

NEVPT2 corrections were built on top of the DMRG configuration interaction (DMRG-CI) calculations, which were performed with the QCMaquis program. The active space was selected based on the single orbital entropies using autoCAS \cite{Stein2019, Bensberg2024b} and a bond dimension of $250$ for the underlying DMRG-CI \cite{White1992, Baiardi2020} calculation. In order to obtain a consistent active space among different sampled geometries, relaxed orbitals were localized using the intrinsic bond orbital (IBO) scheme \cite{Knizia2013} within the PySCF. An active space consisting of 5 Ruthenium $d$-orbitals was found to be appropriate for this system. The full quantum core contained 43/42 occupied and 310/311 virtual alpha/beta orbitals. The DMRG-CI calculation on the selected active space was performed without bond dimension truncation and the optimized wavefunction was used to generate the $n$-RDMs ($n=1-4$) needed for the NEVPT2 correction. Inactive orbitals were canonicalized prior to the NEVPT2 calculation using the generalized Fock operator constructed with the DMRG-CI $1$-RDMs.

\begin{figure}
    \centering
    \includegraphics[width=0.7\textwidth]{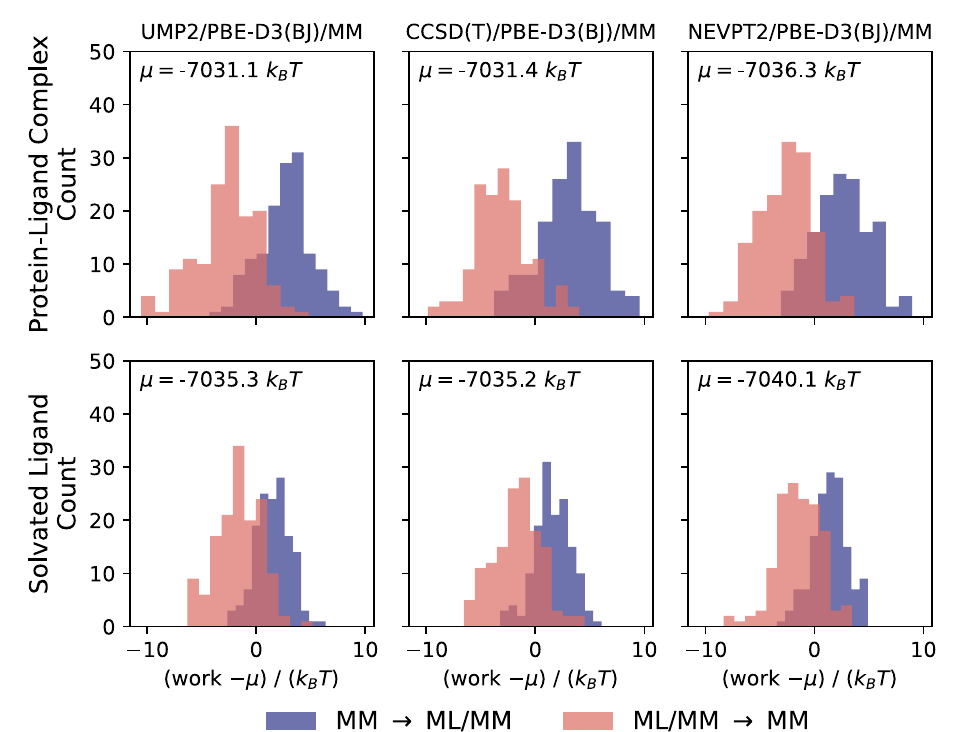}
    \caption{Work distribution for the first (out of six) NEQ simulations for the GRP78-NKP1339 protein-ligand complex, the solvated NKP1339 ligand, and each Huzinaga embedding variant (UMP2/PBE-D3/MM, UCCSD(T)/PBE-D3/MM, and NEVPT2/PBE-D3/MM). The distributions were shifted by their mean $\mu$ for clarity.}
    \label{fig:work-distributions}
\end{figure}

After transfer learning, the QM/QM/MM (UMP2/PBE-D3/MM, NEVPT2/PBE-D3/MM, or UCCSD(T)/PBE-D3/MM) energies, six NEQ switching simulations were performed for the solvated NKP1339 ligand and the solvated GRP78-NKP1339 protein-ligand complex, using the protocol described in Ref.~\cite{Q4Bio-Embedding}. These NEQ simulations provided a distribution of 36 $\Delta G_\mathrm{binding}^\mathrm{ML/MM}$ values. The final value reported in the main article was calculated as its average. The work distributions for the first NEQ simulations of the solvated ligand, protein-ligand complex based on the three Huzinaga embedding variants UMP2/PBE-D3/MM, UCCSD(T)/PBE-D3/MM, and NEVPT2/PBE-D3/MM are shown in Fig.~\ref{fig:work-distributions}. The distributions are narrow and show significant overlap between the switches from the MM to the ML/MM potential energy surface and back. There is no qualitative difference between the distributions obtained with UMP2/PBE-D3/MM, UCCSD(T)/PBE-D3/MM, and NEVPT2/PBE-D3/MM. The narrow distributions and the significant overlap show that the relatively short switching times of $10$\,ps are sufficient to arrive at converged free energy estimates.

\begin{figure}
    \centering
    \begin{subfigure}[b]{0.49\textwidth}
        \vspace{0pt} 
        \centering
        \includegraphics[width=\textwidth]{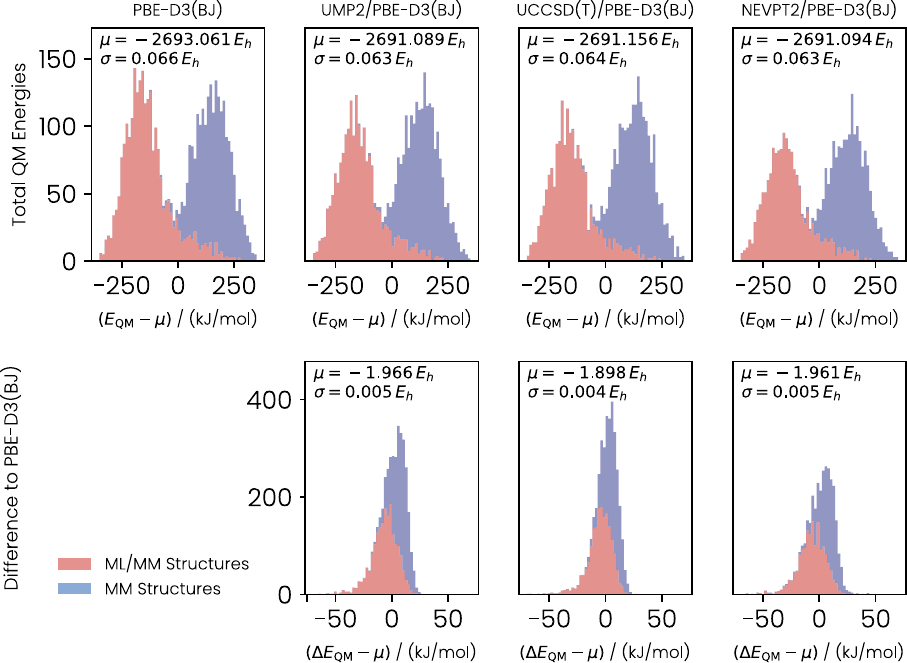}
        \caption{}
        \label{fig:energy-distribution-complex}
  \end{subfigure}
  \begin{subfigure}[b]{0.49\textwidth}
        \vspace{0pt} 
        \centering
        \includegraphics[width=\textwidth]{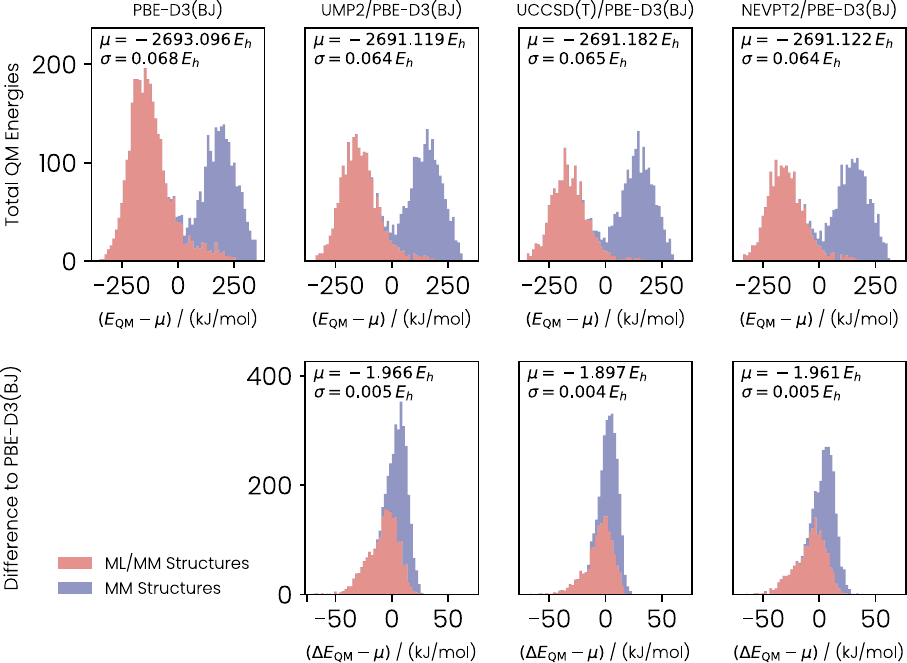}
        \caption{}
        \label{fig:energy-distribution-ligand}
  \end{subfigure}
  \caption{
  Energy distributions obtained for the Huzinaga-based QM/QM/MM embedding, (a) the GRP78/NKP1339 protein-ligand complex, and (b) the solvted NKP1339 ligand. The distributions were shifted by their mean $\mu$, and the standard deviation $\sigma$ is provided for clarity. The distributions for MM and ML/MM structures are stacked, i.e., non-overlapping.
  }
  \label{fig:energy-distributions}
\end{figure}

\sloppy The QM energy distributions for all three Huzinaga-based embedding approaches (UMP2/PBE-D3/MM, UCCSD(T)/PBE-D3/MM, and NEVPT2/PBE-D3/MM) and the ML1 energy model (PBE-D3/MM) are shown in Fig.~\ref{fig:energy-distribution-complex} and Fig.~\ref{fig:energy-distribution-ligand} for the GRP78/NKP1339 protein-ligand complex and NKP1339 solvated ligand, respectively. Furthermore, the differences between QM/QM/MM and the PBE-D3/MM energies are shown. This difference must be effectively learned by the transfer learning approach, elevating the ML1 to the ML2 machine learning potential. The distributions of QM/MM and QM/QM/MM energies is bimodal, because structures from the initial MM force field (which are included in this set) have too short Cl--H distances leading to the right peak (indigo), whereas the DFT computed quantum energies lead to the left peak (red), as discussed in Ref.~\cite{Q4Bio-ML}. The difference between the QM/QM/MM and QM/MM energies is unimodal and narrow because the QM/MM energies already capture a significant portion of the energy.

For the alternative bootstrap embedding-based inner quantum-in-quantum layer, QM/QM/MM calculations were performed using \texttt{QuEmb} \cite{quemb}.
The methodology has been described in detail in previous works \cite{Ye_BEAtom-2019,Ye_MolBE-2019,Ye_CCSDBE-2020,Welborn_BE-2016}. Bootstrap embedding utilizes the mean-field-level description of the full system to generate a set of small-fragment Hamiltonians that are subsequently evaluated using high-level wavefunction methods. The method forms an ensemble of small fragments, based on atomic connectivity, that overlap and span the entire chemical system. Bootstrap embedding uniquely treats the entire quantum region on an equal footing and converges to the full system result with increasing fragment size. This systematic improvability is especially important for near-term quantum devices, since bootstrap embedding provides a fine controllable knob between the necessary number of qubits and target accuracy. As such, it can accommodate the resource limitations of near-term devices and also provides a clear pathway for more accurate results with hardware improvements. The method has been validated on a wide variety of chemical systems in minimal and dense basis sets by demonstrating systematic convergence to full system high-level method energies at reduced cost \cite{Ye_CCSDBE-2020,Q4Bio-Embedding-3b,Tran_BigBasisBE-2024}. Recent extensions to perform bootstrap embedding within the QM/MM framework allow the QM/QM/MM-type embedding calculations necessary for large systems in biochemical simulations \cite{Q4Bio-Embedding-3b}.

\texttt{QuEmb} interfaces with \texttt{PySCF} \cite{Sun2018} to perform the necessary quantum chemistry routines. For bootstrap embedding calculations, mean-field references were obtained with unrestricted Hartree-Fock (UHF) calculations using 
def2-SVP basis set \cite{Ahlrich2005}
and def2-SVP effective core potential for ruthenium \cite{Andrae1990}. L\"{o}wdin localization was used to construct the atom-centered set of orbitals. As introduced in previous work, we generate the entangled interacting bath for each fragment in both the $\alpha$ and $\beta$ densities, using unrestricted CCSD to solve each fragment \cite{Tran_ube-2020}. We omitted chemical potential or density matching, which may lead to further improvements, in this work.
Likewise, we used BE$(1)$ fragmentation, which assigns each non-hydrogen atom and its connected attached hydrogen to separate fragments. 

Raw data generated for this work will be made available on zenodo upon publication of this work.
\fussy

\begin{figure}
    \centering
    \includegraphics[width=0.85\linewidth]{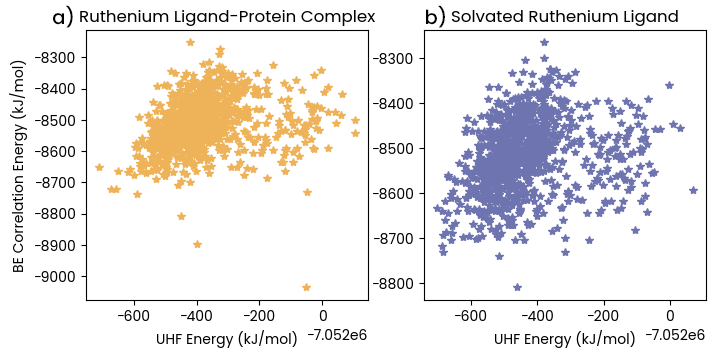}
    \caption{We show the correlation between the bootstrap embedding correlation energies and the UHF energies for each ruthenium ligand-protein complex (a) and solvated ruthenium ligand structure (b). We show that the correlation energies cannot be trivially predicted from the UHF results for either system.}
    \label{fig:be_corr_plot}
\end{figure}

\subsection*{Quantum computing resource estimates}

In the following we give details of how the quantum computing resource estimates of Fig.~\ref{fig:ruth-depth} have been obtained. They are essentially based on five different calculations. First, the active spaces of the quantum core located on the ruthenium drug molecule bound to the target protein need to be set up. This defines the Hamiltonian terms, which enter the resource estimates for quantum phase estimeation. Here we use three methods (qDRIFT, partially randomized, qubitization). Finally, guiding state overlaps need to be computed. 

Active spaces were selected from the set of valence orbitals on the basis of the largest single orbital entropies obtained by autoCAS \cite{Stein2019, Bensberg2024b}. Single orbital entropies were estimated from a low bond dimension ($\chi = 250$) DMRG calculation using QCMaquis \cite{Keller2015}. For each active space, orbitals were ordered using the Fiedler vector of the mutual information \cite{Barcza2011Jan} and the final DMRG calculation was performed to obtain a proxy of the ground state. The bond dimension $\chi=1024$ was found to be sufficient in all cases. Subsequently, we performed a symmetry-shift for the particle number symmetry, as described in Ref.~\cite{Q4Bio-PhaseEstimation}. 
This lowers the sum of weights $\lambda = \sum_{i=1}^L \abs{h_i}$ for the active space Hamiltonians in Pauli representation, $H=\sum_{i=1}^L h_i P_i$ (i.e., after the fermion-to-qubit mapping).

The resource estimation for qDRIFT and the partially randomized method essentially follows the cost analysis done in Ref.~\cite{Q4Bio-PhaseEstimation}. Hamming weight phasing is used for both methods, which for the randomized method requires an estimate on the expected length of commuting terms when sampling. We used an estimated length of about six for all active spaces. Together with the value of $\epsilon,\lambda$ and $\xi$ this is sufficient to arrive at resource estimates for qDRIFT. The partially randomized method is based on the second order Trotter formula and a representation of the Hamiltonian as a weighted sum of Pauli strings. The partitioning of the Hamiltonian terms into deterministic and randomized ones was chosen such that the \textit{total} cost (in terms of Toffolis) is minimized. Additionally, the cost of the partially randomized method depends on the Trotter error, $\epsilon_{\text{Trotter}} = C\delta^2$, where $\delta$ is the Trotter step size. Computing $C$ exactly is as hard as computing the ground state energy, so for larger system sizes an approximation scheme is needed.
In Ref.~\cite{Q4Bio-HPC} exact Trotter errors were computed for the same active spaces of the Ruthenium ligand for up to $16$ spatial orbitals.
We performed a power-law fit of the Trotter error estimates against the $\lambda$ value of the active space Hamiltonians, arriving at $C^{1/2} \approx 1.08\cdot 10^{-4} \lambda^{1.25}$.
Since $\lambda$ is easy to compute even for the largest active spaces studied here, this fit allowed us to estimate $C$ even for larger systems. The resource estimates for the qubitization approach refer to the `double-factorized’ variant \cite{von2021quantum}, and they were done using the \textsc{openfermion} python package \cite{McClean_2020}.

The overlaps were computed by first converging DMRG calculations to have a meaningful proxy for the ground state.
This required going up to bond dimension $1024$. MPS thus obtained was used to construct the guiding states. The lower bond dimension guiding states were obtained by truncating the bond dimension using the singular value decomposition of the density matrix, which maximizes the overlap of the truncated and initial MPS \cite{chan2002}. The Sum-of-Slater guiding states were obtained by sampling the Slater determinants corresponding to the largest coefficients using the sampling-reconstruction of the complete active space (SR-CAS) \cite{boguslawski2011srcas}. Guiding states were normalized in all cases.

\bibliographystyle{unsrtnat}

\end{document}